\documentclass[10pt,amsmath,amssymb,aps,prl,twocolumn]{revtex4-2}
\usepackage{graphicx}
\usepackage{dcolumn}
\usepackage{bm}
\newcommand{\be}{\begin{equation}}
\newcommand{\ee}{\end{equation}}
\newcommand{\ba}{\begin{array}}
\newcommand{\ea}{\end{array}}
\newcommand{\bea}{\begin{eqnarray}}
\newcommand{\eea}{\end{eqnarray}}
\newcommand{\la}{\left\langle}
\newcommand{\ra}{\right\rangle}
\newcommand{\lla}{\left\langle\left\langle}
\newcommand{\rra}{\right\rangle\right\rangle}

\newcommand{\lboro}{\it Nonlinear and Complex Systems Group, \\ Department of Mathematical Sciences, Loughborough University, U.K.}

\begin{document}

\title{Random Networks of Spiking Neurons: \\
Instability in the {\it Xenopus} Tadpole Moto-Neural Pattern}
\author{Carlo Fulvi Mari}
\affiliation{\lboro}
\received{6 August 1999}
\begin{abstract}
A large network of integrate-and-fire neurons is studied analytically when the synaptic weights are independently randomly distributed according to a Gaussian distribution with arbitrary mean and variance. The relevant order parameters are identified, and it is shown that such network is statistically equivalent to an ensemble of independent integrate-and-fire neurons with each input signal given by the sum of a self-interaction deterministic term and a Gaussian colored noise. The model is able to reproduce the quasi-synchronous oscillations, and the dropout of their frequency, of the central nervous system neurons of the swimming {\it Xenopus} tadpole. Predictions from the model are proposed for future experiments. 
\\
\\
\noindent PACS numbers: 87.19.La, 05.45.-a, 87.18.Sn
\end{abstract}

\maketitle

In the recent past the study of neuronal dynamics has been a major topic, both for the importance it has in neuroscience and for the challenge it represents to modern nonlinear mathematics. Particular interest is evoked by the recurrent networks of many neurons, since they are present in many neuronal systems, in mammal cortex as well as in small vertebrates' central nervous system. Specific of the latter is the case of the {\it Xenopus} tadpole embryo, in which the relationship between sensorial stimuli and the consequent behavior of the animal can be studied in detail at the neuronal level \cite{Roberts3,Roberts4}. Pools of reciprocally connected excitatory neurons, distributed bilaterally along the spinal cord, are the elements of the moto-neural circuits of the {\it Xenopus} tadpole that are mainly responsible for the generation of the swimming pattern, consisting of regular oscillations of the body of the tadpole on the horizontal plane. Each pool includes a few hundreds of excitatory neurons receiving inputs from the sensory system, and producing outputs that are sent to the other neurons of the same pool, to pools of inhibitory neurons that act on the symmetrical contra-lateral excitatory pool, and to neurons responsible for muscular activation \cite{Roberts2}. The only inhibitory signals to the neurons of the pool come from outside the pool, mainly caused by activity of the contra-lateral pool. The rhythm of the swim may be not, or at least not only, due to the reciprocal inhibition between the pools on the opposite sides of the spinal cord. Indeed, it has been shown \cite{Soffe} that any excitatory pool is able to present a series of oscillations even if the commissural connections between the two sides of the spinal system are removed and the inhibition is suppressed. The mechanism producing the oscillations in this isolated pool is not fully understood yet. 

In order to investigate on the basic mechanisms of the neuronal activity in recurrent networks, particularly in the case of the {\it Xenopus}, in the present work a large network of $N$ integrate-and-fire (IF) neurons with Gaussian synapses is studied. The membrane potentials are here referred to the reset value $V_{\rm{r}}=0$ and are measured in terms of the threshold potential $V_{\rm{th}}=1$.  Time is measured in units of membrane time constant $\tau=1$.  The equation followed by the membrane potential of neuron $i$ between any two consecutive spikes is 
\be
\frac{dV_{i}}{dt}(t)=-V_{i}(t)+I+\sum_{j\neq i} w_{i j} 
\sum\limits_{m=0}^{\infty}
J(t-T^{m}_{j}), 
\label{IFeq}
\ee 
where $I$ is the external current, and $T^{m}_{j}$ is the instant in which neuron $j$ fires its $m$-th spike. The coefficient $w_{i j}$ is the synaptic weight that multiplies the ionic current density due to the spikes arriving to neuron $i$ from neuron $j$. The weights $\{w_{i j}, i\neq j\}$ are taken to be independently randomly distributed according to a Gaussian probability density function with mean and variance equal to $\mu/N$ and $\sigma^{2}/N$, respectively. Thus, every $w_{i j}$ includes the `sign' of the interaction: positive for an excitatory synapse and negative for an inhibitory one. The weights are quenched and are not required to be symmetric. The function $J(t)$ contains the details of the ionic current density through the synaptic channels, and in all the analytical part of the work its functional form is not relevant. In the numerical study, it is taken as a $\alpha$-function that accounts for the dynamics of the synaptic channels: $J(t)=\alpha^{2} (t-\tau_{a}) \theta(t-\tau_{a}) \exp\{-\alpha (t-\tau_{a})\}$, where also the axonal delay time $\tau_{a}$, for simplicity taken to be the same for all the axonal projections, has been included to simplify the notations.   

Let ${\underline{T}}$ be the matrix of the firing times of all the neurons. Any quantity relative to the network must be a function of ${\underline{T}}$, for example $F({\underline{T}})$. Of particular interest are the quantities that do not depend on the specific realization of the network when the number of neurons is very large (thermodynamic limit). This property (self-averaging) allows one to substitute the calculation of $F$ for a specific realization, usually unfeasible, with the calculation of its average over all the possible realizations of the random network. Normally, one cannot prove {\it a priori} whether the considered quantity self-averages, and one has to calculate also the fluctuation of $F$ across the realizations. 

The average of $F$ across the ensemble of networks is 
\be
\lla F({\underline{T}})\rra =\int d{\underline w}\   
{\cal P}({\underline w}) 
\int d{\underline{T}} \enskip F(\underline{T}) \frac{1}{Z} \delta\left(
{\underline G}({\underline w},{\underline{T}}) \right), 
\label{averageF}
\ee
where the partition function $Z=\int  d{\underline{T}} \enskip \delta\left({\underline G}({\underline w},{\underline{T}}) \right)$ guarantees the correct normalization of the $\delta$-function. The deterministic dynamics of the system is represented by the matrix ${\underline G}({\underline w},{\underline{T}})$, whose generic element gives a relation between two consecutive spikes of a neuron and the spikes previously fired by the other neurons (coming from the IF dynamics, like in \cite{BrCoDes}). For the application of interest in this letter, the uniform initial conditions are adopted here according to which all the neurons are initially stably at membrane rest potential. Thus, for any $n$ and $i$ one has:  
\bea
G^{n}_{i}({\underline w},{\underline{T}})&=&e^{T^{n+1}_{i}} - 
\frac{I}{I-1} e^{T^{n}_{i}} \nonumber \\ &-& \frac{1}{1-I} 
\int_{T^{n}_{i}}^{T^{n+1}_{i}} dt \ e^{t} 
\sum_{j\neq i} w_{i j} \sum_{m} J(t-T^{m}_{j}). \nonumber \\
\eea
With simple appropriate modifications, the mathematical framework here developed can be applied also to many cases in which the neurons of the net are given different initial conditions and different external currents. 

The quantities of interest in this paper can be obtained through the usual techniques based on derivatives of fictitious fields introduced at a later stage during the calculation of the `free-energy' $\lla \ln Z \rra$, and thus the following analysis concentrates on that. 

To calculate the free-energy, the replica trick is used:
\be
\lla \ln Z \rra  =\lim\limits_{r \rightarrow 0} \frac{1}{r} \ln \lla Z^{r} \rra ,
\ee
where the limit is obtained after having calculated the average for integer $r$ and then having prolonged $\lla Z^{r}\rra$ analytically to real exponents. Substituting the $\delta$-functions with their complex Fourier expansions, one has that 
\be
\lla Z^{r}\rra  =\int d{\underline w}\ D{\underline T} Ds \ 
{\cal P}({\underline w}) \exp\left\{\sum\limits_{i,n,\alpha} 
i s_{i}^{n,\alpha} G^{n}_{i}({\underline w},{\underline{T}}^{\alpha})\right\}, 
\ee
being $i$, $n$, $\alpha$ indices of, respectively, neuron, spike, and replica. Through functional integration over one-time and two-times auxiliary functions one can decouple neurons. Due to the absence of dynamic noise and of any form of average over the initial conditions, the replicas are identical also in their kinematics, and this allows for an exact replica-symmetry assumption. Following the symmetry, after the saddle-point evaluation allowed by the extensiveness of the network ($N \rightarrow \infty$), only two order parameters remain in the formulas, both having direct physical meaning as averages over the ensemble of networks. They are: 
\be
\nu(t)=\lla \frac{1}{N} \sum\limits_{i} \sum\limits_{m} J(t-T^{m}_{i})\rra   
\label{nu}
\ee
and
\be
q(t_{1},t_{2}) = 
\lla  \frac{1}{N} \sum\limits_{i} \sum\limits_{m} J(t_{1}-T^{m}_{i}) 
\sum\limits_{n} J(t_{2}-T^{n}_{i})\rra  . 
\ee
Thus, the order parameters `filter' the spike train $\{ T^{n}_{i}, \forall n\}$ fired by neuron $i$ through the current density function $J(\cdot)$. It suggests that the correct smoothing function to analyse the information conveyed by a spike train in real data may be the synaptic function $J$. It can be shown that the arithmetic averages over the neurons, inside the double angle brackets, have vanishing fluctuations in the thermodynamic limit of the mean-field theory (self-averaging); then the order parameter physics make sense also for any single network out of the ensemble. At the level of the single network, the parameter $\nu(t)$ may be seen as an estimate of the average firing rate in the network if the firing rate is measured in a delayed back-time window about $3/\alpha$ long. Actually, $\nu(t)$ weights the preceding spikes according to their synaptic `efficacy' at time $t$. The order parameter $q(t_{1},t_{2})$ is the average auto-correlation of the `filtered' spike trains from the neurons of the network. Another two-times physical order parameter would appear if dynamic noise was present in the network. In fact, both their physical meaning and the analysis of the saddle-point equations show that the two two-times order parameters are equal to each other thanks to the absence of dynamic noise.  

The analysis allows for the interpretation of the mean-field system in terms of an independent IF neuron with total input equal to $I+\hat{\tau}(t)+\mu \, \nu(t)$, being $\hat{\tau}(t)$ a zero-mean Gaussian stochastic process with auto-correlation $\sigma^2 q(t,t')$. Indicating the average over the realizations of the stochastic input with single angle brackets, the order parameters follow the equations 
\be
\nu(t)=\la \sum\limits_{m} J(t-T^{m}) \ra,
\label{nuequiv}
\ee
and
\be
q(t,t')=\la \sum\limits_{m} J(t-T^{m}) \sum\limits_{n} J(t'-T^{n}) \ra,
\label{correquiv}
\ee
where $T^{n}$ is the $n$-th spike time of the equivalent neuron, and the sequence of spiking times is determined by the IF dynamics of the equivalent neuron with input $I+\hat{\tau}(t)+\mu \, \nu(t)$.

Solving analytically this new set of equations, as well as the original saddle-point equations, is a difficult task. However, simulating the equivalent neuron is easy numerically, and finite-size effects are no longer present, as noted by \cite{EissfellerOpper} and \cite{StillerRadons}, since the thermodynamic limit has already been performed. In order to simulate the equivalent system and obtain quantities of interest for the original network, use is made here of the technique introduced by \cite{EissfellerOpper}, where the correlated stochastic input is constructed during the running of a quite large number of independent simulations of the equivalent neuron. In this method, the correlation function is estimated from sample averages, and from that the Gaussian colored signal is created. As suggested by \cite{EissfellerOpper,StillerRadons}, since each one-dimensional simulation is independent from the others, the error on the averages across a set of $M$ trials should scale as $1/\sqrt{M}$. 

An exhaustive analysis in the parameters' space is being worked out and will be presented elsewhere, as well as an analysis of the possibility of substituting the averages over the realizations of the Gaussian noise in Eq. \ref{nuequiv} and Eq. \ref{correquiv} with time averages over a single realization of the noise (ergodicity) that may be useful, e.g., in numerical computations. The rest of this letter concentrates on the particular range of parameters that is of interest for the study of the neural control of the swimming pattern in the simple vertebrate {\it Xenopus} tadpole.

\begin{figure}[t]  
\includegraphics{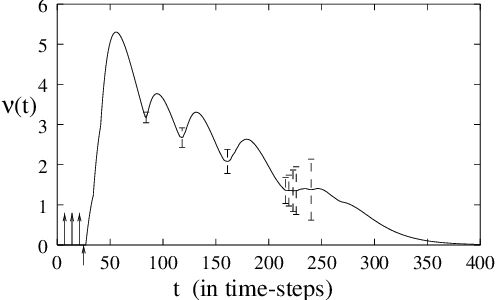}
\caption[]{\footnotesize{Time evolution of the order parameter $\nu(t)$ in the network that mimics a {\it Xenopus} tadpole excitatory pool of motor neurons. The first three arrows indicate the spikes due to the initial external input. The fourth arrow (below the abscissae axis) indicates the instant of the removal of the external source. The bars indicate the standard deviations of the average spike trains [that is, ${\pm \sqrt{q(t,t)-\nu(t) \nu(t)}}$]; they clearly report a progressive synchrony loss. [$\mu=0.9$, $\sigma^{2}=0.001$, $\alpha=5$, $\tau_{a}=0.2$, $I=15$ (only until the 25-th time-step), $M=10000$, time-step=0.01.]}} 
\label{xenopus}
\end{figure}

It has been shown \cite{Soffe} that any excitatory pool surgically isolated from the rest of the tadpole nervous system is able to perform several oscillation cycles after a brief external stimulation that is provided to a large number of neurons inside the pool. It is biologically plausible to assume that the synapses between neurons of the excitatory pool in the {\it Xenopus} spinal system have weights very similar to each other (in the model it means: $\mu>0$ and $\sigma \ll \mu$). In the experiments on the tadpole considered here a brief external input is provided simultaneously to a large number of neurons in a quiescent pool. Thus, the neurons start firing synchronously. The external stimulus is brief, and after its removal the network cannot stay stable. Simulations of the equivalent neuron show that the already present activity decays to zero if $\mu$ is not too large; if $\mu$ is above a certain bound, the activity in the network increases restlessly to saturation. The reason of such instability is discussed later in this letter. For the moment, consider a case with $\mu$ below the aforementioned critical bound. Figure \ref{xenopus} shows the time evolution of the order parameter $\nu(t)$ in a network with $\mu=0.9$ and $\sigma^{2}=0.001$. At the beginning, all the neurons are in a quiescent state, and a constant uniform input current $I$ is applied (0-th time-step). After a while, the external source is abruptly removed, and the network is let to its own dynamic evolution, during which it exhibits a quasi-synchronous activity while the maximal magnitude of the order parameter $\nu(t)$ decays, correspondingly to the also evident decrease in spike frequency. Evident as well is the progressive loss of synchrony (vertical bars; cf. figure caption). The frequency drop continues until the network rests in a quiescent state. Here a case of quite rapid dropout is shown, but larger $\mu$ (still below the bound) can provide slower dropout. The smoothness of the curve near its local minima is an effect of the small variance of the synaptic strengths, that perturbs the synchrony of the spikes, with a consequent smoothing of the average. In complete absence of noise, that is, when all the synapses have equal weight, the curve of $\nu(t)$ presents sharp downward peaks in the local minima (discontinuities of the first time derivative) in delayed correspondence with the spike events. The decrease in synchrony may account for progressive single-neuron spike dropout found in experiments, often interpreted as a spike or synaptic failure. Increasing $\sigma$ causes quicker synchrony loss. These results reproduce the experimental data of frequency dropout \cite{Soffe}. A more quantitative comparison will be presented when appropriate experimental data are available. Meanwhile, direct simulations of the IF network have been carried out: the agreement between theoretical predictions and simulation outputs is very good already for networks with as few as 100 neurons. This also supports the use of the thermodynamic limit as an approximation of the real finite network. 

\begin{figure}[t]  
\includegraphics{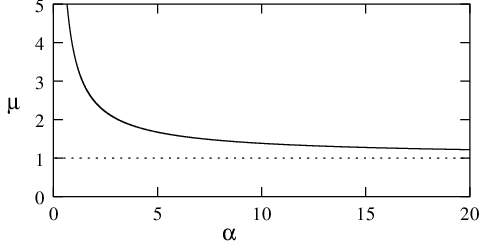}
\caption[]{\footnotesize{Minimum synaptic strength necessary to generate the second spike (cf. text), as a function of the synaptic constant $\alpha$.}} 
\label{mumin}
\end{figure}

The form of instability of self-sustained activity presented by the network in the region of the parameters' space that mimics the tadpole biological values is due to the absence of any synaptic currents' and channels' reset after the generation of a spike, so that they affect the dynamics of the membrane independently of it. It is easily understood how this causes instability in the isolated network: Consider the case of null synaptic variance, for simplicity, and no external input, and assume that all the neurons fire simultaneously the first spike, as an initial condition, for example, so that there are no residual synaptic currents from previous spikes or external sources. Thus, after time $\tau_{a}$, any neuron of the network receives $N$ spikes and starts integrating the synaptic currents. With some analysis of Eq. \ref{IFeq}, it is possible to show that the minimum synaptic strength $\mu$ that allows the membrane potential to reach the threshold depends on $\alpha$ as depicted in Fig. \ref{mumin}. If $\mu$ is below the curve, the neurons do not fire any other spike. On the contrary, if $\mu$ is above the curve the neurons fire at finite time from the arrival of the first spikes; after time $\tau_{a}$, every neuron receives the second wave of spikes, and the new synaptic currents are integrated together with the remains of the currents generated by the first wave, and the neuron fires again. This process is iterated, and the effect of the remaining currents is to decrease monotonically the inter-spike interval, thus driving the network to its maximum firing rate. Simulations of the equivalent neuron seem to indicate that the introduction of a small variance in the synaptic weights lowers the critical bound.

\noindent 

To test the hypothesis about the network instability, numerical solutions have been found introducing in the model the reset of the input currents to zero immediately after every generated spike. The calculations as presented here cannot include this effect, but for small values of $\sigma^{2}$, like in the case of the tadpole modelling, the results are still valid for not too long times. Simulations with this proviso show that when $\mu$ is above the bound, the synaptic variance only destroys synchrony, while the network is in a stable firing state. When $\mu$ is below the critical bound, the network is not able to fire more than one spike after the removal of the external stimulus. One can also obtain stability at low firing-rates substituting the synaptic $\alpha$-function with a function that vanishes at finite time, without a currents' hard reset. If two consecutive spikes do not come at intervals shorter than the time necessary for the synaptic current to vanish, current remains do not cumulate and the network can fire stably for large $\mu$. This latter case would be easily and exactly reproducible in the mathematical framework developed here. 

Since the synaptic currents in the {\it Xenopus} tadpole embryo seem to be well approximated by the $\alpha$-function \cite{Robertscomm}, the results of the present work predict that the mean synaptic strength $\mu$ in the real network is below the critical bound, and the series of oscillations subsequent to the removal of the external stimulus is due to a `surfing' on the remains of synaptic currents until their exhaustion. The present model also predicts a progressive loss of synchrony during the oscillatory transient, due to the small deviation of the synaptic strengths from the mean. A further implication of the present results is that some mechanisms proposed in the past as responsible for the oscillation dropout, like synaptic depression or habituation, are not needed. 

If the predictions about mean synaptic strength, synchrony loss, and single-spike dropout are verified experimentally, the present model may be useful in understanding some neural mechanisms subserving the small vertebrate's motion. With the aid of the mathematical results presented here, other, more general topics are under current study, like the possible presence of several attractors, chaos, and existence of a glassy phase in random networks. 

I thank Professor Paul Bressloff for having brought the spiking neurons' random networks to my attention and for useful discussions. Research supported by UK {\nobreak EPSRC} Grant GR/K86220.

\end{document}